\documentclass{iaus}
\usepackage{graphicx}

\title[Distribution of ALFALFA Galaxies] 
{The Distribution of ALFALFA Galaxies}

\author[Martin]   
{Ann M. Martin$^1$}

\affiliation{$^1$Department of Astronomy, Cornell University,
Ithaca, NY 14853, USA \break email: amartin@astro.cornell.edu}

\pubyear{2007}
\volume{244}  
\pagerange{xxx-xxx}
\date{?? and in revised form ??}
\jname{Proceedings Dark Galaxies and Lost Baryons}
\editors{J.I. Davies \& M.D. Disney, eds.}
\begin{document}

\maketitle

\begin{abstract}
The ALFALFA$^{2}$ \footnotetext{$^{2}$ The Arecibo Observatory is part of the National Astronomy and Ionosphere Center which is operated by Cornell University under a cooperative agreement with the National Science Foundation.}blind HI survey will enable a census of the distribution of gas-rich galaxies in the local Universe. Sensitive to an HI mass of 10$^{7}$ solar masses at the distance of the Virgo cluster, ALFALFA will probe the smallest objects locally and provide a new consideration of near-field cosmology. Additionally, with a larger, cosmologically significant sample volume and wider bandwidth than previous blind surveys, a much larger number of detections in each mass bin is possible, with adequate angular resolution to eliminate the need for extensive follow-up observations. This increased sensitivity will greatly enhance the utility of cosmological probles 
in HI. ALFALFA will eventually measure the correlation function of HI selected galaxies in a large local volume. The larger sample and volume size of the ALFALFA dataset will also robustly measure the HI mass function (HIMF). Here, we present the preliminary results on the distribution of local gas-rich galaxies from a first ALFALFA catalog covering 540 deg$^{2}$.

\end{abstract}

\firstsection 
\section{First ALFALFA Catalog}
A catalog of the Northern Virgo Cluster region has recently been published (\cite{Giovanelli07}), including 730 detections of galaxies and high velocity clouds (HVCs) in the Milky Way or its periphery over 132 deg$^{2}$ (approximately 1.7$\%$ of the total ALFALFA sky). Here, we present the mass and redshift characteristics of an extended version of the Northern Virgo Cluster region catalog with greater extent in right ascension, 07$^{h}$30$^{m}$ $<$ R.A. $<$ 16$^{h}$30$^{m}$, for a total of 540 deg$^{2}$. Objects are detected over a bandwidth range -1,600 km s$^{-1}$ $<$ cz $<$ 18,000 km s$^{-1}$. There are 94 HIPASS detections in this region, while there are $\sim$2700 ALFALFA detections, 2.5$\%$ HVCs and 97.5$\%$ galaxies. Of these galaxies, 3$\%$ have no as yet identifiable optical counterpart, but these are typically associated with other structures (Giovanelli $\&$ Kent, these proceedings).

\section{Mass and Redshift Distribution}
\begin{figure}
 \includegraphics[height=1.75in,width=5in]{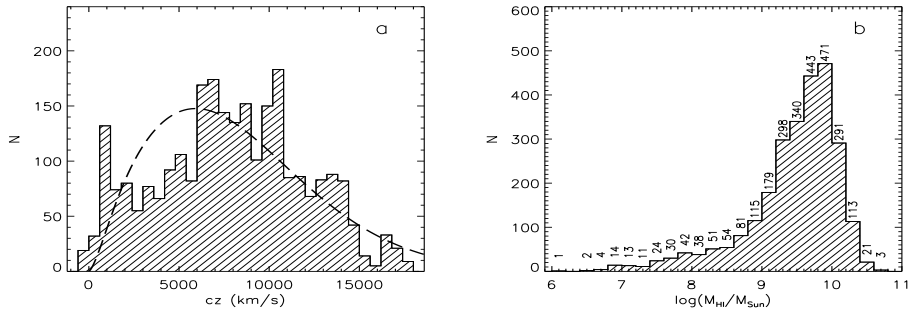}
  \caption{Histograms of preliminary HI candidate detections. HVCs are excluded. (a) Recessional velocity in km s$^{-1}$. Dashed line: expected redshift distribution for a homogeneous distribution of galaxies and an HIMF following that of \cite{Zwaan03}.  The presence of large-scale structure in this region is clearly exhibited; note especially the overdensity near 1,000 km s$^{-1}$ due to the Virgo Cluster. (b) Derived HI mass, in logarithmic units of M$_{\odot}$. }\label{fig:hists}
\end{figure}

The median redshift of the catalog is 7,700 km s$^{-1}$, and the redshift distribution of the full sample (see Figure~\ref{fig:hists}(a)) reflects the known local large scale structure, including the Virgo cluster, the void behind it, the A1367-Coma supercluster at 7,000 km s$^{-1}$, and a third, more distant overdensity at 13,000 km s$^{-1}$. The impact of radio frequency interference leads to apparent underdensities in redshift space. Very strong RFI is associated with the San Juan FAA radar operating at 1350 MHz; this leads in particular to a loss of data near 15,000 km s$^{-1}$, and samples will not be complete in this region.

ALFALFA distances are estimated using a flow model (Masters 2005; \cite{Saintonge07}) to 3,000 km s$^{-1}$, after which a pure Hubble flow distance is used. HI mass is calculated via $M_{HI} = 2.36 \times 10^{5} D^{2} F_{c}$ where $D$ is the distance in Mpc and $F_{c}$ is the integrated line flux in Jy km s$^{-1}$. ALFALFA probes both extremes of the distribution more completely than previous blind HI surveys; the histogram in Figure~\ref{fig:hists}(b) shows the mass distribution estimated for this extended catalog. 16$\%$ of the objects in this catalog are more massive than 10$^{10}$ M$_{\odot}$, and 5$\%$ have masses between 10$^{6}$ and 10$^{8}$ M$_{\odot}$. The highest mass objects detected by ALFALFA are of particular interest; ALFALFA probes a cosmologically significant volume of the Universe at $z = 0$, and will help to pin down both ends of the HIMF. Future cosmological studies targeted toward the redshift evolution of the HIMF will be biased in favor of such high mass objects, and will rely on the ALFALFA detection of large numbers of galaxies with masses greater than 10$^{10}$ M$_{\odot}$. Blended emission from pairs or groups are likely to be the explanation for many of the highest-mass detections. 

\section{Future Work \& Conclusions}
As the survey proceeds, this sensitivity will allow the HI mass function (HIMF) to be probed at the problematic low-mass end, where current estimates vary by an order of magnitude near 10$^{6}$ M$_{\odot}$. ALFALFA's ability to probe the low-mass end of the HIMF depends, however, on distance uncertainties. While 21-cm observations are ideal for large-scale surveys because they come with a redshift, it is difficult to determine a distance to an HI-selected galaxy without extensive followup in other regimes. Fractional distance errors are greatest for low recessional velocities, where we are most likely to detect the lowest-mass objects; this problem is exacerbated in the high-density (and high peculiar velocity) environment of the Virgo cluster.  For this reason, an HIMF of ALFALFA galaxies awaits further completion of the survey.

\begin{acknowledgments}
This work has been supported by NSF grants AST--0307661,
AST--0435697, AST--0607007, by the Brinson Foundation, and by the National Defense NDSEG Fellowship. 
\end{acknowledgments}

\end{document}